# Superconducting TSV contact for cryoelectronic devices

Ivan Filippov [1], Alexandr Anikanov [1], Aleksandr Rykov [1], Alexander Mumlyakov [1], Maksim Shibalov [1], Igor Trofimov [1], Nikolay Porokhov [1], Yuriy Anufriev [1], and Michael Tarkhov [1]

[1]Institute of Nanotechnology of Microelectronics of the Russian Academy of Sciences, Nagatinskaya 16a-11, Moscow 115487, Russia

**Abstarct**

This work focuses on the fabrication of niobium through silicon vias (TSV) superconductors interconnects. The effect of supercycle of sequential oxidation and chemical etching process on the through-etch wall quality was investigated. It was experimentally shown that the use of supercycle in the fabrication process leads to significant improvement of the TSV wall quality and removal of the defect type - scallops. After 12 times repetitions of supercycles a dissipative bonding of superconducting strips on the front and back side of the sample is observed. The critical current density of such coupling is $5 \times 10^4$ A/cm$^2$. The critical ratio of substrate thickness to hole diameter at which electrical coupling is formed is 3 : 1.

1. **Introduction**

Interest in superconducting quantum computing has increased dramatically in recent years, mostly due to the potential of increasing the computing processing power of such devices in connection with classical processors [1]. Qubit processors have demonstrated the basis of quantum error correction protocols, elementary quantum algorithms and simulations [2]. Universal gate operations are performed with 99.9% accuracy for single qubits and 99.5% accuracy for two-qubit gates. The use of optimised parametric amplification allows non-destructive qubit measurements with more than 99% confidence [3]. Coherence time of qubits is constantly increasing and has reached 150 μs [4]. At the same time, the high-speed classical control electronics required for real-time feedback also continue to evolve rapidly.

Designing and fabricating large scale superconducting circuits with addressable, low noise cross-talk switching for all circuit elements is a complex task. Superconducting qubits are sensitive to fabrication defects, which limits the yield and reproducibility of the final device. Both aspects require optimised device and process design.

Modern superconducting applications and devices require an increase in the integration density to combine them with silicon CMOS technology, which is mostly used as the controlling part. To achieve this coupling, through-silicon-vias integrated packaging techniques are being developed, which allow multiple layers of crystals to be logically interconnected. A similar technology has long been used in CMOS devices [5].

There are different technological approaches for forming layer-to-layer interconnects. One approach suggests fabricating 300 μm deep channels using aluminium with an aspect ratio of 6:1, to interconnect a CMOS crystal with qubit layer [6]. Materials like Al, TiN, Nb and Nb alloys (NbN, NbTi and NbTiN) investigates to fabricate supeconducting TSV with different types shapes of channels. Best result was achieved using Al/TiN layers with single, sharp and hysteresis free transition that was measured at 1.27 K [7]. Another approach, the use of hybrid flip-chip technology via indium bumps to control signal routing in two-dimensional highly coherent circuits, has demonstrated the potential for further deep integration of quantum devices [8-9]. The through-hole geometry with sloped walls for uniform deposition of aluminium, which provides a reliable connection between the superconducting quantum integrated circuit and the control board, showed zero resistance below the critical temperature [10].

The integration of superconducting qubits with superconducting TSVs with high aspect ratio and 200 μm depth is demonstrated in [11-12]. Interconnects are used for baseband control and high-precision microwave qubit readout, the ability to fabricate qubits directly on the chip surface has also been shown.

The main technological challenges in implementing superconducting interconnections are to eliminate wall roughness from etching of high aspect ratio silicon structures and conformal deposition of metal inside these channels without breaking the conductive layer.

In this paper, we present for the first time, a method to fabricate niobium-based superconducting interconnectors formed by a high aspect ratio etching (Bosch process), thermal oxidation and chemical etching "supercycle" followed by conformal Nb sputtering. The combination of these techniques allows us to achieve ultra-smooth TSV channel walls with further conformal niobium layer deposition to form a superconducting interconnect in an interposer layer. With this approach, we have fabricated a TSV chip that can be integrated into future qubit assemblies and other cryoelectronic devices, where 2 or more chips levels can be used. A prototype of the fabricated structure is shown in Figure 1.

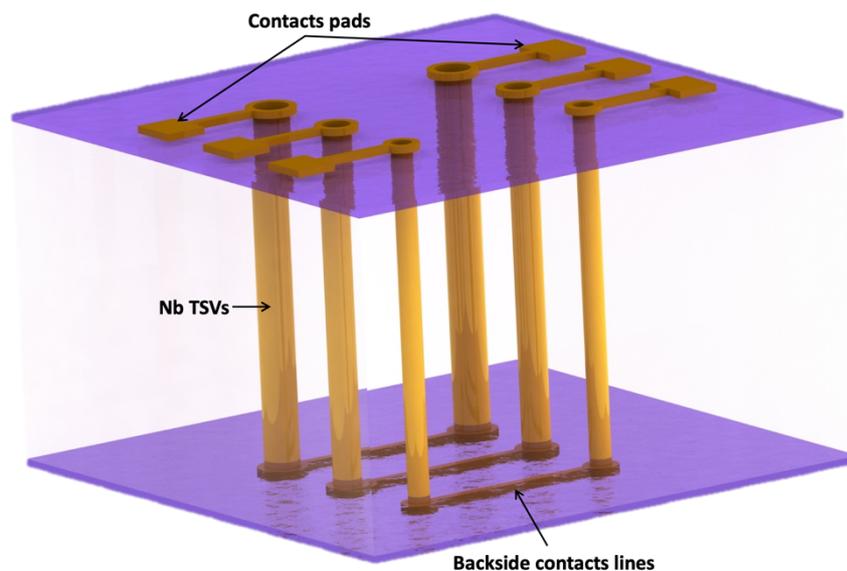

Figure 1 - 3D sketch of a crystal realized with the proposed technology, including: 2 crystal layers and a TSV superconducting interlayer.

## 2. Fabrication technology

For the fabrication of the superconducting TSVs, 460-μm-thick and double-side polished silicon wafers are used. The fabrication process started with a 3.8-μm-thick thermal silicon oxide forming operation, which would later be used as an etch mask for the Bosch process. SiO2 layer, was patterned by laser lithography using 5 μm AZ4999 photoresist and then etched by anisotropic plasma chemical etching process in C4F8 gas. Main parameters of process are: 5 sccm flow for 35 minutes with ICP power 1500 W and DC bias more than 90 V. The etching rate in this mode is 110 nm per minute. Holes with sizes 80, 100 and 150 microns are formed in thermal silicon oxide. After the etching operation, the surface was cleaned from photoresist residues using oxygen plasma. For deep silicon etching, a three-step Bosch process is used, involving deposition of a protective polymer layer in plasma C4F8 gas, opening the bottom of the formed structure with high DC bias (above -100V) in plasma SF6 gas and the final isotropic etching of silicon using the same gas. Preliminary to the Bosch process, the substrate surface was stripping in oxygen-argon plasma mixture for 1 minute with DC bias near -320 volts. The duration of each step in Bosch process, including the polymer deposition, bottom breakthrough and etching steps ranged from 2 to 4 seconds and were determined experimentally for used thicknesses of silicon wafers. The last 100 etch cycles were carried out using a LF generator at the bottom opening step to avoid notching the effect, with an DC bias -150 volts. Also, due to the increased aspect ratio, the etch cycle time was increased by 0.5 seconds in the last 100 μm of etching. Simultaneously with the TSV channels in the Bosch process, thorough lithographic marks were formed, for subsequent reverse crystal alignment. A schematic of the main steps of fabrication process shown in Figure 2.

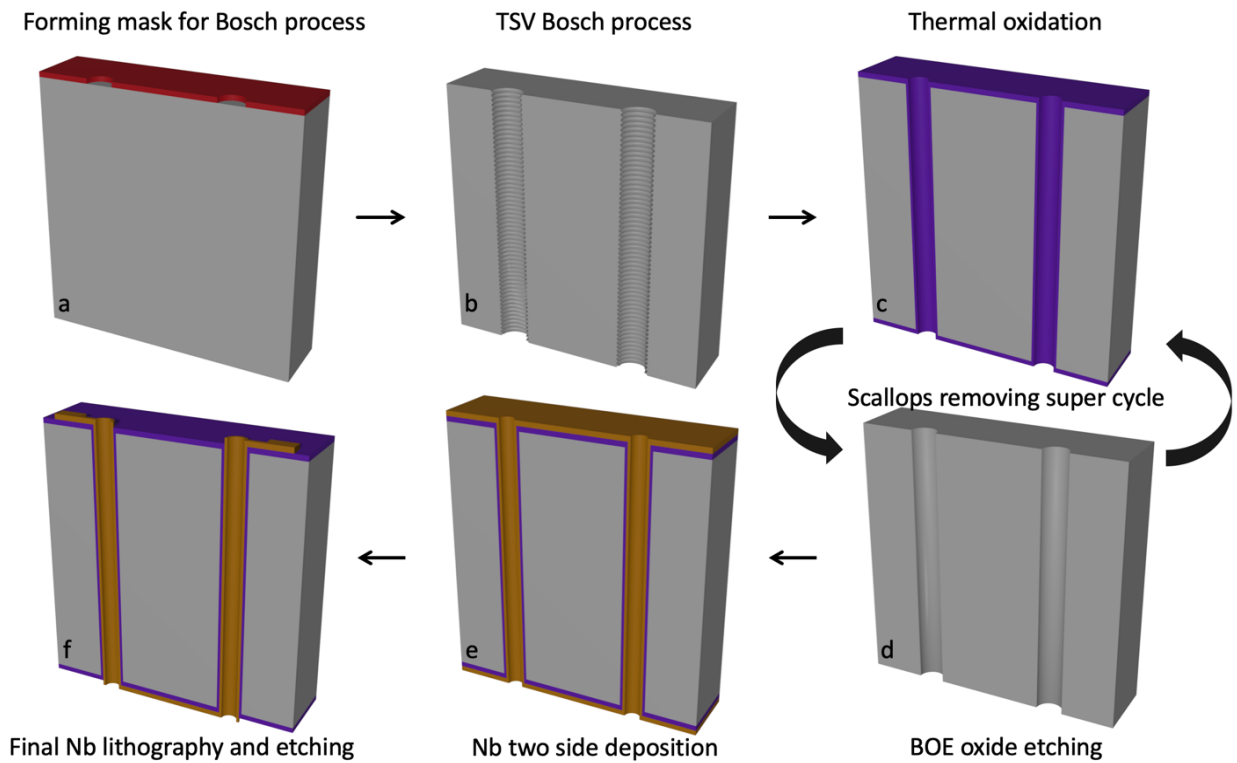

Figure 2 – Graphical presentation of the main steps in the process route for the fabrication of TSV interconnects: a – Preparing mask for TSV etching, that is the window in SiO2 layer by laser photolithography and plasma etched anisotropic process ; b - Bosch process 460 μm deep and removal of top and bottom silicon oxide in BOE solution; c and d - Re-oxidation and removing of silicon oxide several times from TSV channels; e - Deposition of 200 nm thick niobium on both sides; f - Lithography and etching of niobium layers on both sides.

The formation of wall roughness in the Bosch process, called scallops [13], is well known to occur due to alternating passivation followed by an isotropic etching process, whose size depends on the length of one cycle of silicon deep etching. For our process the typical size of scallops is 300 - 500 nm. By depositing a 200 nm thick niobium layer using a magnetron process, the conformality of the layer is also affected by the geometric shape of the walls and the connection between the layers is lost.

After the Bosch process, the structures were cleaned in oxygen plasma from the passivation residue and the mask silicon oxide layer was removed in BOE solution. Further steps in the fabrication of TSVs can be combined in a "supercycle", which includes sequential thermal oxidation at high temperatures in water vapour and chemical etching of the formed oxide to completely remove the scallops on the walls. To achieve the lowest roughness in the TSV channels, three wafers were experimentally compared with 3, 9 and 12 repetitions of supercycle. The supercycle consists of the following technological operations: chemical cleaning of the silicon wafer after the Bosch process which includes treatment in Peroxymonosulfuric acid solution

($H_2O_2/H_2SO_4$ = 3/7, t = 10 min, T = 120 °C) and peroxide-ammonia solution ($H_2O/H_2O_2/NH_4OH$ = 5/1/1, t = 10 min, T = 75 °C); thermal oxidation in water vapour at T = 1000 °C. The thickness of the obtained silicon oxide is 300 nm; etching of thermal $SiO_2$ in BOE buffer solution (t = 20 min). The process of etching the oxide in BOE buffer etchant (ammonium hydrodifluoride $NH_4FHF$) is based on the following reaction: $SiO_2 + 6HF \rightarrow H_2SiF_6 + 2H_2O$, where $H_2SiF_6$ is water soluble. A schematic representation of the supercycle is shown in Figure 3a.

This sequence was repeated for all wafers. Before deposition of the superconducting layer of niobium, a chemical treatment cycle and thermal oxidation in water vapour at 100 nm was also performed for the wafers to form a dielectric barrier layer. The result of 200 nm thick niobium film deposition before (Figure 3b) and after 12 times of repeating of supercycle is shown in Figure 3c.

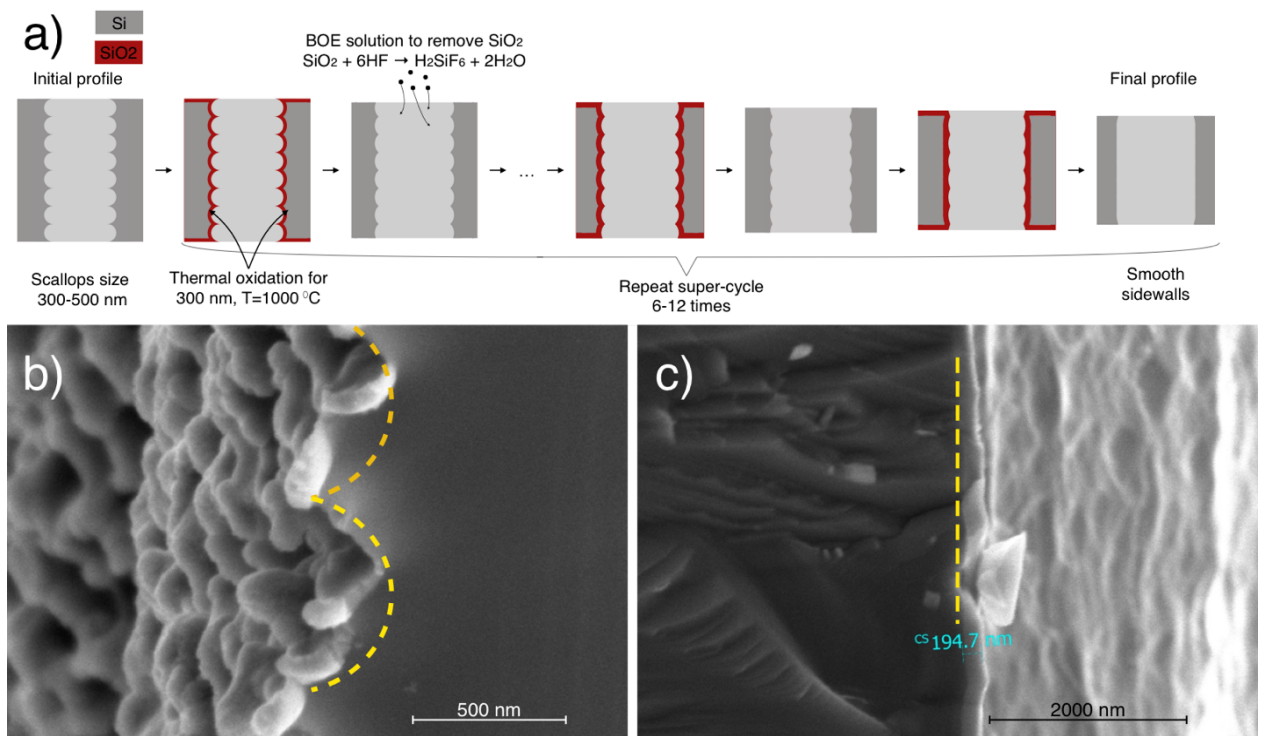

Figure 3 - (a) Schematic of the supercycle process; (b) SEM image of cross-sectional view of TSVs channel walls with deposited niobium before supercycle; (c) SEM image of cross-sectional view of TSVs channel walls with 200 nm thick deposited niobium s after 12 supercycle runs.

The two-sided coating of niobium film, on a 100 mm wafer, was obtained by standard reactive magnetron sputtering. The deposition process was carried out at room temperature from a 99.999% pure niobium target using argon gas with a DC magnetron power of 5 W/cm² and a deposition rate of 14 nm per minute under these conditions. The magnetron chamber was equipped with an ion-source with energies up to 2.5 keV to activate the substrate surface. Films were deposited sequentially on the both sides of the substrate with a vacuum break between sputtering processes.

After deposition of the superconducting layer, a contact pattern was formed on both sides using 950-nm thickness S1809 photoresist. The image of the resulting contacts is shown in Figure 4. To form the superconducting strips, the niobium was etched in $SF_6/Ar$ plasma mixture with a 2:1 ratio, at RF and ICP source powers of 20 W and 800 W, respectively.

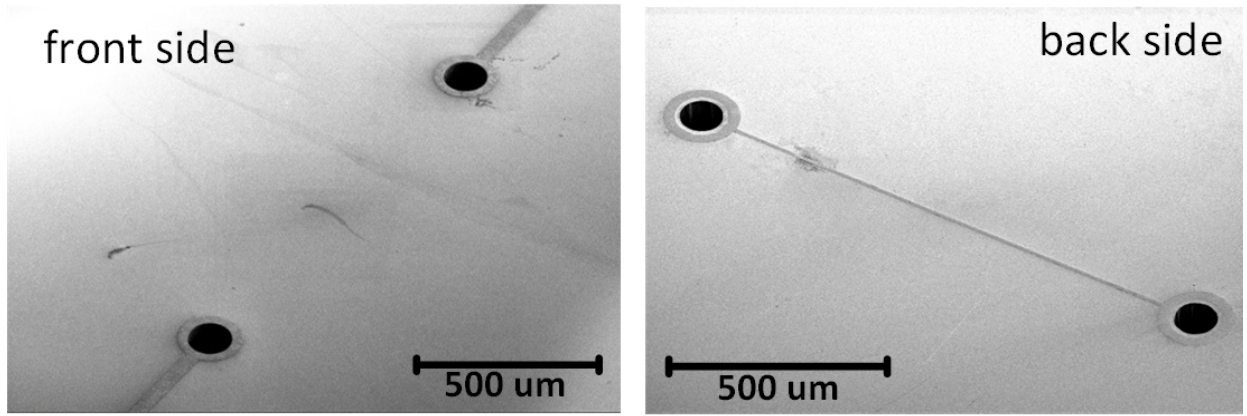

Figure 4 - SEM images of fabricated TSV interconnections and Nb supercondacting stripes on front and back sides.

For the best cover of the TSV channel walls with niobium during its deposition into high aspect ratio structures, the confocal orientation of the magnetron sputter target with a substrate holder was used. Figure 5a shows a schematic of a magnetron unit with ion assist capability during the deposition of metal films. The angle between the substrate holder and the normal to the magnetron is 35 degrees. In situ assisted ion beam argon activation with an energy of up to 2.5 keV, which ensures high adhesion and quality of the niobium film in the through-etch area was used. During niobium film deposition, the substrate holder was rotated at 60 rpm. Figures 5b and 5c schematically show the niobium coating of channel walls in a confocal film deposition configuration. In the middle of the TSV channel, "defective" overlayering is formed due two deposition processes, that are made on the front and backside of the substrate.

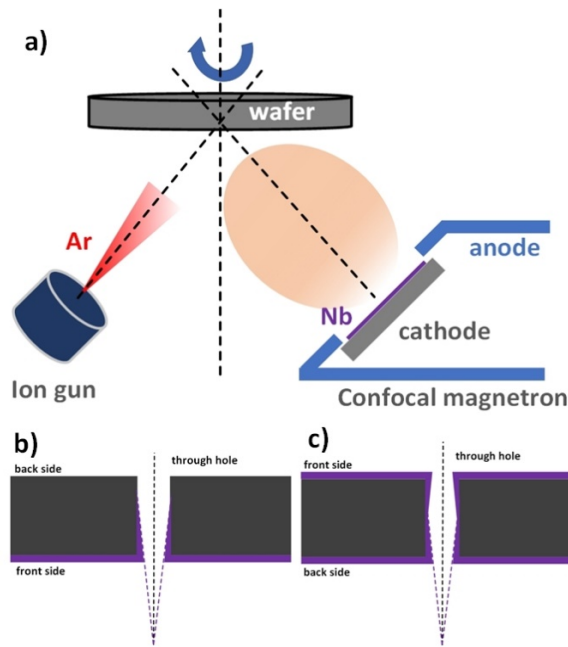

Figure 5. a) Schematic representation of a confocal sputtering magnetron. b) Front side sputtering process. (c) Back side sputtering process.

## 3. Experimental results

In the first stage, the resistance of the structures was measured at room temperature to determine the quality of the Nb TSV interconnections. TSVs with hole diameters of 80 and 100 μm had resistances much larger than calculated. Samples with a diameter of 150 μm matched the calculated resistivity, indicating that the contact between the layers was formed.

The temperature dependence of the contact resistance with TSV was measured in a Gifford-McMahon closed-cycle cryostat. The temperature was measured with a temperature controller on calibrated diode thermometer. Resistance measurement was carried out with a Keithley 2460 precision source-meter at a bias current of 1 μA. Our possible assumption of such resistance values for 80 μm and 100 μm holes is related to the formation of native oxide layer after first Nb deposition in the middle of the channel where overlap of the two metal layers occurs.

Figure 6 shows the temperature dependences of the normalized resistivity for samples with 3, 9 and 12 supercycles, respectively. The transition of niobium contacts occurs at 9K, which corresponds to bulk niobium. However, due to the influence of etch wall defects, the residual resistance has rather large values up to 9 supercycle. We experimentally show that 12 supercycles lead to superconductive coupling between the electrodes on the front and back side of the structure.

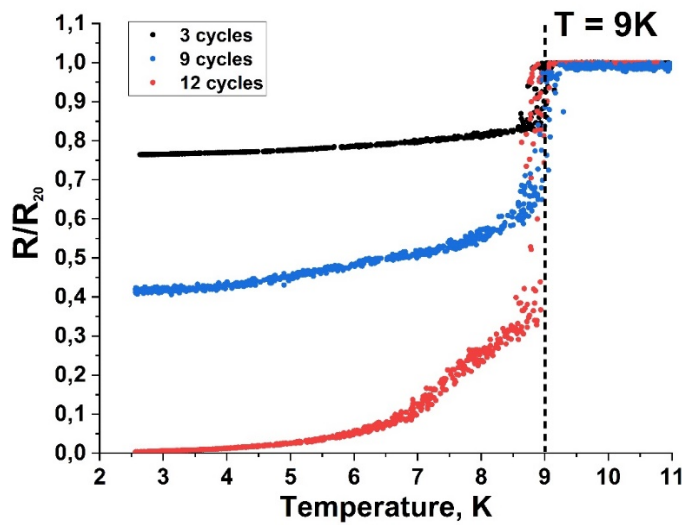

Figure 6. Dependence of normalised resistivity on temperature for a sample with 3, 9 and 12 supercycles, respectively.

However, it can be seen from the temperature dependence of the resistance that the influence of the "defective" film area leads to a delayed transition caused by partial oxidation of the film during the change of deposition side.

Also, current density measurements were made at the lowest achievable temperature of 2.4K, which was $5 \times 10^4$ A/cm$^2$. The current densities obtained are not record values, but for most applications the current density is not important value as it absolute value in dissipative current flow.

## 4. Conclusion

High-aspect, superconducting 460 μm deep niobium interconnects fabricated by a combination of Bosch process methods and super cycles of thermal oxidation and chemical etching were demonstrated. This method allows superconducting layer-based TSV interconnects to be integrated into standard CMOS manufacturing processes, which will improve packing density for future advanced cryo-electronics applications. Electrical characterisation of the fabricated crystals showed a critical current density of $5 \times 10^4$ A/cm$^2$ at 2.4 K.

The proposed technology of using a sequential oxidation and etching process makes it possible to substantially eliminate the influence of scallop-type wall defects. It is important to note that the use of thinner substrates will result in significantly improved performance. It has been empirically shown that the critical ratio of substrate thickness to hole diameter is ~3 : 1, at which point the TSV channel walls are conformally coated.

## Acknowledgements

This work is dedicated in memory of our leader, colleague and friend Elena Valentinovna Zenova.